\documentclass[reprint,aps,prx,showkeys]{revtex4-2}

\usepackage{amsmath,amssymb}
\usepackage{color}
\usepackage{graphicx}
\usepackage{bm}
\usepackage{cancel}
\usepackage{mathtools}
\usepackage{ulem}
\usepackage[shortlabels]{enumitem}

\usepackage[colorinlistoftodos]{todonotes}
\DeclareMathSymbol{\R}{\mathalpha}{AMSb}{"52}

\begin{document}


\title{A General Model for Linearly Polarized Optical Vector Beams}


\author{J. M. Nichols$^1$ and F. Bucholtz$^2$}
\address{$^1$
U.S. Naval Research Laboratory, 4555 Overlook Ave. SW., Washington D.C. 20375\\
}
 
\address{$^2$Amentum Technology, Inc., 4800 Westfields Blvd, Chantilly, VA 20151\\
}

\date{\today}

\begin{abstract}
We propose an approach for deriving a broad class of propagation models for inhomogeneously, linearly polarized ``vector'' beams.  Our formulation leverages a complex scalar potential along with an appropriately constructed Lagrangian energy density. Importantly, we show that polarization inhomogeneities can be included by simple addition of a spatially dependent polarization angle to the complex potential phase.  Thus, phase and polarization are seen to be equivalent from an energy perspective.  As part of our development, we also show how the complex scalar potential arises naturally when considering polarization angle as a field symmetry during construction of the Lagrangian.  We further show that the definition of linear momentum density in terms of the complex potential holds a distinct advantage over the conventional definition for inhomogeneously polarized beams.  
\end{abstract}


\maketitle



\section{Introduction}

In 1953 Green and Wolf \cite{Wolf:53} made the observation that a complex {\it scalar} potential could be used to represent an electromagnetic {\it vector} field.  Defining this potential $V({\bf x},t)$ as a general function of both space and time, they then introduced an associated Lagrangian energy density
\begin{align}
    \mathcal{L}&=\frac{\epsilon_0}{2}\left[\dot{V}({\bf x},t)\dot{V}^*({\bf x},t)-\nabla V({\bf x},t)\cdot \nabla V^*({\bf x},t)\right]
    \label{eqn:LagrangianDensityWolf}
\end{align}
where the overdot denotes time derivative, $\nabla(\cdot)\equiv \{\partial_x(\cdot),\partial_y(\cdot),\partial_z(\cdot)\}$, the operator $(\cdot)^*$ takes the complex conjugate, and $\epsilon_0$ is the vacuum permittivity.  The authors went on to show that in a homogeneous, isotropic medium, $V({\bf x},t)$ obeyed the familiar scalar wave equation,
\begin{align}
    \nabla V({\bf x},t)-\frac{n^2({\bf x})}{c^2}\ddot{V}({\bf x},t)&=0,
    \label{eqn:wave}
\end{align}
frequently used to model the evolution of a particular component of the electric field, $E({\bf x},t)\in {\bf E}({\bf x},t)\equiv\{E_x({\bf x},t),E_y({\bf x},t),E_z({\bf x},t)\}$.  Setting $V({\bf x},t)=E({\bf x},t)$, and solving (\ref{eqn:wave}) for each component independently is in fact the basis of scalar wave optics (see \cite{Goodman:68}, Chapter 3).

However, the real value of $V({\bf x},t)$ is that it can in principle be used to model the full vector ${\bf E}({\bf x},t)$ and not simply a scalar component.  The attraction of such an approach is the ability to capture the effects of coupling among vector components while working with simpler scalar equations.   

Over the years a number of efforts have been aimed toward this objective.  One of the earliest can be attributed to Whittaker \cite{Whittaker:04} who showed that one could represent both the electric and magnetic field vectors as derivatives of two appropriately chosen scalar potentials. This basic idea was later built upon by Pattanayak and Agrawal \cite{Pattanayak:80} who showed that for a linearly polarized beam, only one of the Whittaker potentials is required to model the beam, a result that is consistent with the findings presented here.

Of most relevance to this paper is the aforementioned works of Green and Wolf \cite{Wolf:53}, along with Wolf \cite{Wolf:59}, and Roman \cite{Roman:59}.  This trio of papers established expressions for the flow of energy and momentum in optical fields based on complex scalar potentials and we will leverage these expressions here.  Notably, Berry \cite{Berry:09} made the subsequent claim that a scalar potential could not be used to represent interference between waves with different, general polarization states and that a second potential would be required (e.g., as in \cite{Whittaker:04}).  For a general elliptical state of polarization, this is indeed true.  However, for linearly polarized light, we have found the single scalar potential sufficient as will be shown. 

In this work we present a complex scalar potential that facilitates the modeling of beams with inhomogeneous polarization distributions, or ``vector'' beams.  In general, these beams cannot be modeled by a superposition of scalar solutions to Eqn. (\ref{eqn:wave}) as such an approach cannot capture certain interactions among the vector components of the electric field.  The potential we describe is capable of doing so and provides a simple recipe for deriving vector beam models.  In fact, we will use this potential to  generalize our previously obtained ``paraxial'' vector beam model \cite{Nichols:22}.

We proceed by briefly reviewing the paraxial vector beam model development of \cite{Nichols:22}.  We then show how an appropriate complex scalar potential, in combination with a corresponding paraxial Lagrangian density, can be used to directly obtain the vector beam model via the principle of stationary action.  In doing so we demonstrate that one can view the construction of $V({\bf x},t)$ as the introduction of an optical field symmetry in the Lagrangian energy density togther with a higher-order ``symmetry breaking'' term.  We further show that the definition of linear momentum density offered in \cite{Wolf:53} in terms of the vector potential correctly captures the influence of the polarization gradient whereas the conventional definition does not.

With these results in mind, we can then immediately see how to generalize the paraxial vector beam model.  We create a more general complex scalar potential and insert into the Lagrangian density (\ref{eqn:LagrangianDensityWolf}). The more general vector beam models are then easily obtained via the principle of stationary action.  This new model includes known, homogeneously polarized beam models and our previously derived vector beam model as subsets.

\section{Background: Paraxial Vector Beams \label{background}}

In prior work \cite{Nichols:22,Nichols:24} we studied the dynamics of a linaerly polarized, monochromatic beam propagating through free-space ($n({\bf x})=1$), primarily in the $\hat{z}$ direction.  In this case the electric field vector can be written ${\bf E}({\bf x},t)=\tilde{E}(\vec{x},z)e^{-i(k_0z-\omega t)}$ where $k_0$ is the wavenumber and $\omega=c k_0$ is the temporal frequency and $c$ the speed of light.  The associated complex electric field amplitude vector $\tilde{\bf E}(\vec{x},z)$ is a function of the transverse coordinates $\vec{x}$ and propagation distance $z$. The components are easily shown to be governed by the free space paraxial vector wave equation \cite{Nichols:22}
\begin{align}
\left[-i2k_0\partial_z+\nabla_X^2\right]\{\tilde{E}_x,\tilde{E}_y,\tilde{E}_z\}(\vec{x},z)
\label{eqn:PWE}
\end{align}
where $\nabla_X\equiv \{\partial_x(\cdot),\partial_y(\cdot)\}$ is the transverse gradient.  In accordance with Maxwell's equations the electric field amplitudes must also obey the constraint $\nabla\cdot \tilde{E}(\vec{x},z)=0$ which, in the paraxial approximation, mandates $\tilde{E}_z({\bf x})=ik_0^{-1}\nabla_X\cdot\{\tilde{E}_x({\bf x}),\tilde{E}_y({\bf x})\}$. Thus, there are only two unknowns that must be solved for in (\ref{eqn:PWE}).  

\subsection{Solutions to the Paraxial Wave Equation}

Common practice obtains these transverse electric field amplitudes via the Huygens-Fresnel integrals
\begin{align}
    \tilde{E}({\bf x})&
    =\left(\frac{ik_0}{2\pi z}\right)^{\frac{1}{2}}
    \int \tilde{E}({\bf x}_0) e^{-\frac{ik_0\left((x-x_0)^2+(y-y_0)^2\right)}{2z}}dx_0dy_0
    \label{eqn:HFpx}
\end{align}
where ${\bf x}_0\equiv \{x_0,y_0,0\}$ and $\tilde{E}({\bf x}_0)$ is the scalar transverse electric field amplitude at $z=0$.  For a beam comprised of only a signal non-zero vector component, Eqn. (\ref{eqn:HFpx}) is the exact solution to (\ref{eqn:PWE}). In fact, even when both transverse components are non-zero, if the beam is homogeneously polarized, applying (\ref{eqn:HFpx}) to both transverse coordinates {\it independently} and superimposing the results still gives the exact solution to (\ref{eqn:PWE}).  

To see why, we recognize that rotating our initial coordinate system by a spatially constant polarization angle transforms the two-dimensional problem back into the scalar problem.  That is to say, a homogeneously linearly polarized, transverse vector beam is effectively a scalar beam for which (\ref{eqn:HFpx}) is the exact solution.  Keeping this in mind, take an arbitrary linearly polarized electric field $\tilde{E}_x({\bf x}_0)=\tilde{E}({\bf x}_0)\cos(\gamma),~\tilde{E}_y({\bf x}_0)=\tilde{E}({\bf x}_0)\sin(\gamma)$. 
 In the rotated coordinate system we solve for $\tilde{E}({\bf x})$ exactly using (\ref{eqn:PWE}) and then simply multiply the result by $\cos(\gamma),~\sin(\gamma)$ respectively to form the solution in the original coordinates.  Because the polarization angle is constant (is not part of the integrand in \ref{eqn:HFpx}), this is no different than the aforementioned approach of performing the HF integrals in $\hat{x},~\hat{y}$ respectively and superimposing the result.  Hence this latter approach is also exact in the homogeneously polarized case. 

However, this is not the case for a spatially inhomogeneous polarization distribution where both electric field components are required in the description (i.e., no global coordinate transformation can reduce Eqn. \ref{eqn:PWE} to a scalar problem). 
 Moreover, these components will in general be coupled so that applying (\ref{eqn:HFpx}) to both transverse components independently (and then superimposing the results) no longer yields the general solution of (\ref{eqn:PWE}).  
 
 To further understand these dynamics, substitute the transverse vector electric field model
\begin{align}
&\{\tilde{E}_X(\vec{x},z),\tilde{E}_Y(\vec{x},z)\}=\nonumber \\
&\qquad\rho^{1/2}(\vec{x},z) e^{-i\phi(\vec{x},z)}\left\{
\cos{\gamma(\vec{x},z)},\sin{\gamma(\vec{x},z)}\right\}
\label{eqn:efield}
\end{align}
with magnitude $\rho^{1/2}$, dynamical phase $\phi$, and linear polarization angle $\gamma$ into (\ref{eqn:PWE}).  After simplification, one obtains \cite{Nichols:24}
\begin{subequations}
    \begin{align}    \partial_z\rho(\vec{x},z)+\nabla_X\cdot\left[\rho(\vec{x},z)\vec{v}(\vec{x},z)+\rho(\vec{x},z)\vec{\Omega}(\vec{x},z)\right]&=0
    \label{eqn:continuity}
    \end{align}
    \begin{align}
    \frac{D \vec{v}(\vec{x},z)}{Dz}&=-\left(\vec{\Omega}(\vec{x},z)\cdot\nabla_X\right)\vec{\Omega}(\vec{x},z)\nonumber \\
    &\qquad\qquad+\frac{1}{2k_0^2}\nabla_X\left(\frac{\nabla_X^2\rho^{1/2}(\vec{x},z)}{\rho^{1/2}(\vec{x},z)}\right)  
    \label{eqn:velocity}
    \end{align}
    \begin{align}
    \frac{D\vec{\Omega}(\vec{x},z)}{Dz}+\left(\vec{\Omega}(\vec{x},z)\cdot\nabla_X\right)\vec{v}(\vec{x},z)=0
    \label{eqn:vorticity}
    \end{align}
    \label{eqn:model}
    \end{subequations}
where 
\begin{align}
    \vec{v}(\vec{x},z)=k_0^{-1}\nabla_X\phi(\vec{x},z) \nonumber \\
    \vec{\Omega}(\vec{x},z)\equiv k_0^{-1}\nabla_X\gamma(\vec{x},z)
    \label{eqn:vDef}
\end{align}
and where $D(\cdot)/Dz\equiv \partial_z(\cdot)+(\vec{v}\cdot\nabla)(\cdot)$ denotes the material derivative.
Absent a polarization gradient ($\vec{\Omega}(\vec{x},z)=0$), the model (\ref{eqn:model}) is solved exactly by the electric field components obtained via (\ref{eqn:HFpx}).  In other words, for a homogeneously polarized beam the components of the paraxial vector field (\ref{eqn:efield}) can indeed be propagated independently via the Huygens-Fresnel integrals.  
However, if the polarization angle is spatially dependent, treating the components independently will fail to capture the beam behavior and one must instead solve the coupled partial differential equations (\ref{eqn:model}).
For example, a polarization gradient can be designed such that the beam centroid follows a parabolic path through free-space \cite{Nichols:22}. Performing the diffraction integrals (\ref{eqn:HFpx}) predicts no such effect and the beam the centroid is predicted to remain unchanged.  Carrying out the experiment as in \cite{Nichols:24} shows that indeed the beam centroid follows the path predicted by (\ref{eqn:model}).  The physical mechanism for this transverse acceleration is an exchange of the two types of transverse momentum (see section \ref{sec:momentum}) captured inside the divergence operator in Eqn. (\ref{eqn:continuity}), an effect that cannot be captured by independent applications of Eqn. (\ref{eqn:HFpx}). 

\subsection{Stationary Action Solutions for Vector Beams \label{sec:action}}

An alternative derivation of the model (\ref{eqn:model}) proceeds from the principle of stationary action.  This approach suggests finding those values of the system variables $u^{(i)}$ that minimize variations in 
\begin{align}
\mathcal{A}&=\int_{\R^4} \mathcal{L}\left(u^{(i)},\nabla u^{(i)},{\bf x},t\right) d{\bf x} dt
\end{align}
where $\mathcal{L}(u^{(i)},\nabla u^{(i)},{\bf x},t)$ is the Lagrangian energy density. The solution $\mathcal{L}$ that renders the action stationary on the domain defined by our spatio-temporal coordinates can be shown through the calculus of variations (see e.g., \cite{Stone:09} Eqn. 1.81 or \cite{Berra:23} section 5.1), to obey 
\begin{align}
\frac{\partial\mathcal{L}}{\partial u^{(i)}}-\nabla\cdot\left[\frac{\partial\mathcal{L}}{\partial(\nabla u^{(i)})}\right]-\partial_t \left(\frac{\partial \mathcal{L}}{\partial_t u^{(i)}}\right)=0.
\label{eqn:lagrangeeqns}
\end{align}   

Consider first the free-space, ``paraxial''  Lagrangian energy density for the complex electric field amplitude $\tilde{E}=\rho^{1/2}\exp(i\phi)$,
\begin{align}
    \mathcal{L}&=-\frac{i\epsilon_0k_0}{2}\left(\tilde{E}^*\dot{\tilde{E}}-\tilde{E}\dot{\tilde{E^*}}\right)-\epsilon_0\nabla_X\tilde{E}^*\cdot\nabla_X\tilde{E}\nonumber \\
    &\textrm{or}\nonumber \\
    \mathcal{L}&=-\epsilon_0k_0\rho\partial_z\phi-\frac{\epsilon_0}{2}\rho|\nabla_X\phi|^2
    -\frac{\epsilon_0|\nabla_X\rho|^2}{8\rho},
    \label{eqn:paraxLagrangian}
\end{align}
Note that in this monochromatic model, time has been effectively removed from the problem as the amplitudes depend only on the three spatial coordinates, hence the Lagrangian density is in units $J/m^3$. 
 The overdot for the amplitudes in this case represent derivatives w.r.t. $z$.  
 
 Using the Lagrangian (\ref{eqn:paraxLagrangian}) and the Euler-Lagrange equations (\ref{eqn:lagrangeeqns}) with $u^{(1)}\equiv \phi,~u^{(2)}\equiv \rho$ then yields the homogeneously polarized beam model, that is, Eqns. (\ref{eqn:model}) but with $\vec{\Omega}=0$. Specifically, taking the variation w.r.t. $\phi$ yields
\begin{align}    \partial_z\rho(\vec{x},z)+\nabla_X\cdot\left[\rho(\vec{x},z)\vec{v}(\vec{x},z)\right]=0
\label{eqn:paraxcont}
\end{align}
while the variation in $\rho$ yields 
\begin{align}
    k_0\partial_z\phi(\vec{x},z)+\frac{1}{2}|\nabla_X\phi(\vec{x},z)|^2=\frac{1}{2}\frac{\nabla_X\rho^{1/2}(\vec{x},z)}{\rho^{1/2}(\vec{x},z)}.
    \label{eqn:paraxeikonal}
\end{align}
Taking the transverse gradient of this expression then yields (\ref{eqn:velocity}). We again point out that the solution of these equations matches the well known results of scalar beam theory.

Now, let us consider again the inhomogeneously polarized case.  We will now show that the extension to the full vector beam model simply involves the transformation $\phi\rightarrow \phi+\gamma$ in the Lagrangian density.  For this modified expression, taking the variation w.r.t. either $u^{(1)}\equiv \phi$ or $u^{(2)}\equiv \gamma$ yields exactly (\ref{eqn:continuity}).  On the other hand, taking the varation w.r.t. $u^{(3)}\equiv \rho$ yields
 \begin{align}
     k_0\frac{D\gamma(\vec{x},z)}{Dz}&+k_0\partial_z\phi(\vec{x},z)+\frac{1}{2}|\nabla_X\phi(\vec{x},z)|^2\nonumber \\
     &+\frac{1}{2}|\nabla_X\gamma(\vec{x},z)|^2
     =\frac{1}{2}\frac{\nabla_X\rho^{1/2}(\vec{x},z)}{\rho^{1/2}(\vec{x},z)}
     \label{eqn:bigeikonal}
 \end{align}

 We can recognize that the total derivative  of polarization angle does not change on propagation, i.e., $D\gamma/Dz=0$.  This is simply a statement that, as an intrinsic property of light, polarization angle is simply ``following'' the local optical path \cite{Nichols:22}.  
 
 Interestingly, one can also understand the transformation underlying (\ref{eqn:bigeikonal}) as a ``local gauge'' that produces, to leading order, a field symmetry \cite{Karatas:90}. 
 The total derivative of polarization angle can then be set equal to zero as a direct consequence of Noether's Theorem (see \ref{sec:derivation}).  That same analysis predicts that the quantity $\int\rho\gamma d\vec{x}$ is conserved on propagation, an intuitive result given that total intensity is conserved and polarization angle is not changing along the local optical path.
 
 The end result is that the expression (\ref{eqn:bigeikonal}) is actually two expressions
 \begin{subequations}
 \begin{align}
     k_0\partial_z\phi(\vec{x},z)&+\frac{1}{2}|\nabla_X\phi(\vec{x},z)|^2+\frac{1}{2}|\nabla_X\gamma(\vec{x},z)|^2\nonumber \\
     &=\frac{1}{2}\frac{\nabla_X\rho^{1/2}(\vec{x},z)}{\rho^{1/2}(\vec{x},z)}
     \label{eqn:phaseeqn}
     \end{align}
     \begin{align}
         \frac{D\gamma(\vec{x},z)}{Dz}&=0
     \label{eqn:totalderivgamma}
     \end{align}
     \label{eqn:velocityvorticity}
 \end{subequations} 
 where the higher order term $|\nabla_X\gamma|^2/2$ ``breaks'' the aforementioned symmetry and can actually be designed to alter the beam path (see again \cite{Nichols:24}).
 Taking the transverse gradient of both equations (\ref{eqn:velocityvorticity}) then yields Eqs.   (\ref{eqn:velocity}) and (\ref{eqn:vorticity}) respectively.  In other words, simply adding the polarization angle to the overall phase in the paraxial Lagrangian density yields the same equations of motion as substituting the full vector electric field into the paraxial wave equation.  The transformation $\phi\rightarrow \phi+\gamma$ is creating a complex scalar amplitude $\tilde{V}=\tilde{E}\exp(i\gamma)$ of precisely the type permitted by the results of \cite{Wolf:53}.

{\it We thus arrive at the important result that the complex scalar potential
\begin{align}
    V({\bf x},t)&=\rho^{1/2}(\vec{x},z)e^{-i[k_0z+\phi(\vec{x},z)+\gamma(\vec{x},z)-\omega t]}
    \label{eqn:paraxscalar}
\end{align}
can be used to model propagation of a monochromatic, linearly polarized, \underline{vector} electric field ${\bf E}({\bf x},t)$ with a spatially dependent polarization angle in the paraxial regime.}  Alternatively stated, using 
\begin{align}
    \tilde{V}(\vec{x},z)=\rho^{1/2}e^{-i[\phi(\vec{x},z)+\gamma(\vec{x},z)]}
\end{align}
in place of the scalar electric field in the complex paraxial Lagrangian (\ref{eqn:paraxLagrangian}) yields the same model (\ref{eqn:model}) as we obtained by substituting the full vector electric field into (\ref{eqn:PWE}).  The resulting system of equations is appropriate for describing the evolution of magnitude, phase, and linear polarization angle in paraxial vector beams.

\subsection{Consideration of Momentum Density \label{sec:momentum}}

From this paraxial example we can also see an important departure from conventional optics in the representation of momentum density.
Defining the transverse electric field vector
\begin{align}
&\{E_x,E_y\}\equiv \vec{E}(\vec{x},z)\nonumber \\
&~~~=
\rho^{1/2}e^{-i(\phi(\vec{x},z)+k_0z-\omega t)} \{\cos(\gamma(\vec{x},z),\sin(\gamma(\vec{x},z)\}
\end{align}
the full, paraxial field becomes  \cite{Bekshaev:07b}
\begin{align}
    {\bf E}({\bf x},t)&=\left\{E_x,E_y,\frac{i}{k_0}\nabla_X\cdot\vec{E}(\vec{x},z)\right\}\nonumber \\
    {\bf H}({\bf x},t)&=\frac{1}{\epsilon_0 c}\left\{-E_y,E_x,\frac{i}{k_0}\nabla_X\cdot\{-E_y,E_x\}\right\}
\end{align}
where the $\hat{z}$ component is determined by enforcing the constraint $\nabla\cdot{\bf E}({\bf x},t)=0$.  The conventional momentum density is then given by 
\begin{align} 
    {\bf P}_C&=c{\bf E}({\bf x},t)\times {\bf H}^*({\bf x},t)\nonumber \\
    &=\left\{k_0^{-1}\rho\nabla_X\phi,\rho\right\}.
    \label{eqn:classical}
\end{align}
The result yields the known definition of transverse optical momentum, $k_0^{-1}\rho\nabla_X\phi$.  Notably, the polarization angle gradient is missing from the definition.  The transverse momentum density for the polarization gradient beam can be obtained directly from the energy conservation equation (\ref{eqn:continuity}) as ${\bf P}_V=k_0^{-1}\rho(\nabla_X\phi+\nabla_X\gamma)$.  In fact, in an earlier work \cite{Nichols:24} we demonstrated that Eqns. (\ref{eqn:model}) could be combined to yield the conservation statements
\begin{align}
\frac{\partial\rho}{\partial z}&+\nabla_X\cdot{\bf P}_V=0\nonumber \\
\frac{\partial {\bf P}_V}{\partial z}&+\nabla_X\cdot\left[{\bf P}_V\otimes \frac{1}{\rho}{\bf P}_V\right]=-\nabla_X\cdot {\bf T}\nonumber \\
{\bf T}&\equiv -\frac{\rho}{4k_0^2}\left(\vphantom{\vec{\Omega}}\nabla_X\otimes\nabla_X\log\rho\right),
\label{eqn:Cauchy3}
\end{align}
the middle expression being the local conservation of transverse momentum.  The polarization angle gradient is therefore an integral part of the transverse momentum density, despite being absent from the conventional result (\ref{eqn:classical}).

Keeping these results in mind, we note that Green and Wolf offered an alternative definition of momentum density in terms of the scalar potential \cite{Wolf:53}
\begin{align}
    {\bf P}_V&=-\frac{1}{2c}\left[\dot{V}^*({\bf x},t)\nabla V({\bf x},t)+\dot{V}({\bf x},t)\nabla V^*({\bf x},t)\right].
    \label{eqn:momdensity}
\end{align}
For the monochromatic, paraxial beam this expression yields for the momentum density
\begin{align}
    {\bf P}_V&=\left\{k_0^{-1}\rho(\nabla_X\phi+\nabla_X\gamma),\rho\right\}
\end{align}
which is precisely the desired result.
{\it Thus, the definition of linear momentum density (\ref{eqn:momdensity}) 
correctly captures the influence of a spatial gradient in linear polarization whereas the conventional definition (\ref{eqn:classical}) does not}.  

From (\ref{eqn:Cauchy3}) it is clear that the momentum contribution from the polarization-gradient is essential to demonstrating momentum conservation in vector beams.  The influence of this additional momentum term was unambiguously demonstrated in experiment as part of a ``momentum exchange'' that accelerates the beam centroid \cite{Nichols:24}.

\section{Proposed Scalar Potential \label{sec:density}}

With the results of the previous sections in mind we can see immediately how to construct a general complex scalar potential appropriate for modeling linearly, inhomogeneously polarized optical fields in an isotropic medium characterized by real-valued refractive index $n({\bf x})$.  To do so we define 
\begin{align}
    V({\bf x},t)&=\rho^{1/2}({\bf x},t)e^{-i\alpha({\bf x},t)}
    \label{eqn:potential}
\end{align}
where $\alpha({\bf x},t)=\Phi({\bf x},t)+\gamma({\bf x},t)$ includes both phase {\it and} polarization, just as in the paraxial case.

Defining the operator
\begin{align}
    {\bm\partial}(\cdot)\equiv \left\{i\nabla(\cdot),\frac{n({\bf x})}{c}\partial_t(\cdot)\right\}
\end{align}
the Lagrangian energy density in $J/m^3$ (note the time operator is normalized by $c$ and hence is converted to a distance) can then be written in several compact forms, all equivalent to (\ref{eqn:LagrangianDensityWolf}):
\begin{align}
\mathcal{L}(V;x,y,z,t)&=\frac{\epsilon_0}{2}\left\{{\bm\partial}V\cdot{\bm\partial}V^*\right\}\nonumber \\
\textrm{or}\nonumber \\
\mathcal{L}(\rho,\alpha;x,y,z,t)&=\frac{\epsilon_0}{2}\rho\left\{{\bm \partial}\alpha \cdot {\bm \partial}\alpha+\frac{{\bm \partial}\rho \cdot {\bm \partial}\rho}{4\rho^2}\right\}\nonumber \\
\textrm{or}\nonumber \\
\mathcal{L}(\beta,\alpha;x,y,z,t)&=\frac{\epsilon_0}{2}e^{2\beta}\left\{{\vphantom{\frac{1}{2}}\bm \partial}\alpha \cdot {\bm \partial}\alpha+{\bm \partial}\beta \cdot {\bm\partial}\beta\right\}
\label{eqn:Lagrange}
\end{align}
where, in the final formulation, we simply used the alternate representation of the amplitude, $\beta\equiv \log(\rho^{1/2})$ (In fact, for a monochromatic beam this final expression is the same as Eqn. 3.22 in \cite{Wolf:53}).

The values of $\rho(x,y,z,t),~\alpha(x,y,z,t)$ that minimize the action associated with (\ref{eqn:Lagrange}) are then found via (\ref{eqn:lagrangeeqns}) to obey
\begin{subequations}
\begin{align}
    -\frac{n^2}{c^2}\partial_t\left[\rho\partial_t\alpha\right]+\nabla\cdot\left[\rho\nabla\alpha\right]&=0
    \label{eqn:Poynting}
    \end{align}
    \begin{align}
    -\frac{n^2}{2c^2}\left(\partial_t\alpha\right)^2+\frac{1}{2}\nabla\alpha\cdot\nabla\alpha&=\frac{1}{2}\frac{\nabla^2\rho^{1/2}}{\rho^{1/2}}-\frac{n^2}{2c^2}\frac{\partial_{tt}\rho^{1/2}}{\rho^{1/2}}.
    \label{eqn:eikonalGen}
\end{align}
    \label{eqn:genmodel}
\end{subequations}
The first of these equations is the result of taking variations in $\mathcal{A}$ w.r.t. $u^{(1)}=\alpha$ and represents the well-known conservation of energy, sometimes referred to as Poynting's Theorem \cite{Paganin:98}.
The second expression governs the phase and is the result of taking variations in $\mathcal{A}$ w.r.t. $u^{(2)}=\rho$.  The gradient in this expression is seen to govern the optical path just as it did in the paraxial case.  Likewise, we can see that the optical path in this more general case is also influenced by the polarization gradient.   

The model (\ref{eqn:genmodel}) is entirely consistent with the scalar form of the wave equation (\ref{eqn:wave}) as proved in \cite{Wolf:53}.  This is confirmed by substituting (\ref{eqn:potential}) into (\ref{eqn:wave}).  After setting both the real and imaginary portions of the result equal to zero, the result is (\ref{eqn:genmodel}).  In fact, (\ref{eqn:genmodel}) matches Eqns. 3.13 and 3.14 in \cite{Wolf:53} but with the critical difference being the presence of a spatially dependent polarization angle in (\ref{eqn:genmodel}).

A typical case considers a monochromatic electric field model with no spatial dependence in polarization. In this case, one sets $\alpha({\bf x},t)=(\Phi({\bf x})+\gamma-\omega t)$ and $\rho({\bf x})$ is a function of space only.  With these assumptions, and setting $n=1$ (free space), Eqns. (\ref{eqn:genmodel}) recovers standard the scalar optics model satisfying the Helmholtz Eqn. (see \cite{Saleh:91} eqn. 2.3-2).  For paraxial, homogeneously polarized beams set $\Phi(\vec{x},z)=\phi(\vec{x},z)+k_0 z$, $\gamma=const$, followed by application of the slowly varying envelope approximation.  The resulting model gives the same solution as that found using the familiar Huygens-Fresnel diffraction integral of scalar, paraxial optics (Eq. \ref{eqn:HFpx}).  

However, the real utility of the (\ref{eqn:genmodel}) comes in modeling vector beams whereby, for example, $\alpha({\bf x})=\Phi({\bf x})+\gamma({\bf x})-\omega t$.  Presuming $\Phi({\bf x})=\phi({\bf x})+k_0 z$ and applying the slowly varying envelope approximation in $\hat{z}$ gives the model of \cite{Nichols:22} which has been subsequently explored in \cite{Nichols:24}. 
  
As a more general example, consider 
$\alpha({\bf x},t)=\Phi({\bf x},t)+\gamma({\bf x},t)-\omega t$ and apply the slowly varying envelope approximation $|\partial_{tt} {\bf E}|\ll \omega |\partial_t {\bf E}|$ (i.e., a quasi-monochromatic beam).  The end result of this approximation is that $\partial_t\phi,~\partial_t\gamma\ll \omega$ so that terms involving products of these quantities or higher order temporal derivatives vanish.  Simplifying the model (\ref{eqn:genmodel}) then yields
\begin{subequations}
    \begin{align}    \partial_t\rho({\bf x},t)+\nabla\cdot\left[\rho({\bf x},t)\vec{v}({\bf x},t)+\rho({\bf x},t)\vec{\Omega}({\bf x},t)\right]&=0
    \label{eqn:continuity2}
    \end{align}
    \begin{align}
    \frac{D \vec{v}({\bf x},t)}{Dt}&=-\left(\vec{\Omega}({\bf x},t)\cdot\nabla\right)\vec{\Omega}({\bf x},t)\nonumber \\
    &\qquad\qquad+\frac{c^2}{2k_0^2}\nabla\left(\frac{\nabla^2\rho^{1/2}({\bf x},t)}{\rho^{1/2}({\bf x},t)}\right)  
    \label{eqn:velocity2}
    \end{align}
    \begin{align}
    \frac{D\vec{\Omega}({\bf x},t)}{Dt}+\left(\vec{\Omega}({\bf x},t)\cdot\nabla\right)\vec{v}({\bf x},t)=0
    \label{eqn:vorticity2}
    \end{align}
    \label{eqn:model2}
    \end{subequations}
where
\begin{align}
    \vec{v}&\equiv \frac{c}{n^2 k}\nabla\phi\nonumber \\
    \vec{\Omega}&\equiv \frac{c}{n^2 k}\nabla\gamma.
    \label{eqn:veldefs}
\end{align}
The model (\ref{eqn:model2}) is the generalization of (\ref{eqn:model}) where the spatial dependence on the propagation direction $z$ has been replaced with dependence in time while the transverse gradient operators become full, three-dimensional gradients.  The ``velocity'' $\vec{v}$ and polarization gradient terms now have the units $m/s$ (as opposed to being dimensionless quantities in Eqns. \ref{eqn:model}). 

The model (\ref{eqn:model2}) is in fact a new transport model for vector beam propagation. 
 Still other models, obtained by applying different assumptions, can also be easily derived from (\ref{eqn:genmodel}).

\section{Summary}

We have proposed a complex scalar potential as a complete representation for linearly polarized beams with spatially inhomogeneous polarization distributions.  Remarkably, such a potential can be constructed by simply adding the polarization angle to the phase term found in previously derived potentials \cite{Wolf:53}.  In combination with an appropriate Lagrangian density, we were then able to derive the paraxial vector beam model presented previously in \cite{Nichols:24}.  Based on this approach, we then used the new complex potential to generalize the model to include non-paraxial vector beams with time-dependent amplitude and phase.

Importantly, we also showed that the complex scalar potential representation led to the correct expression for linear momentum density.  On the other hand, the conventional momentum definition fails entirely to capture the contribution of the polarization angle distribution.  This result will therefore perhaps be useful in resolving ongoing discussions on how to interpret optical momentum.  

The notion that dynamical phase $\phi$ and polarization $\gamma$  contribute equally to an overall phase term $\alpha$ is not entirely surprising.  It has long been known that a geometric Pantcharatnam-Berry phase is in some cases added to the traditional dynamical phase in describing beam propagation \cite{Berry:87}.  It is also known that the generation of a linear polarization gradient results in the creation of a geometric phase (see our prior work \cite{Nichols:22}).  Thus, one could reasonably conclude that a spatially-dependent polarization angle and a phase are describing closely related physics. 

We believe this to be the first work, however, that shows these quantities are equal contributors to the system Lagrangian energy density.  Our interpretation is that while $\phi$ represents the phase common to both transverse electric field components (i.e., the dynamic phase), spatially-dependent polarization can be appropriately viewed as a phase between electric field components.  This is precisely why the effects predicted and observed in \cite{Nichols:24} can {\it only} be described with a vector solution to the wave equation (as opposed to scalar diffraction theory).

Lastly, we were able to represent the transformation from a scalar to a vector beam model as a field ``symmetry''.  Thus, we are afforded the results of Noether's Theorem which we used to provide additional justification for separating the beams' dynamical phase and polarization angle in the governing equations.  This analysis also shows that the product of intensity and polarization angle is a conserved quantity in beam propagation which, to our knowledge, is also a new result.

\section{Acknowledgments}
The authors would like to acknowledge support of the Office of Naval Research Code 33 under grant N0001422WX01660 

\iftrue

\appendix

\section{Linear Polarization and Noether's Theorem}\label{sec:derivation}

In section (\ref{sec:action}) We provided justification for the 
separation of the expression (\ref{eqn:bigeikonal}) into the two separate mathematical statements (\ref{eqn:velocityvorticity}), one for each ``phase'' variable $\phi,~\gamma$.  Here we provide an alternative, mathematical argument based on field theory. 

Our basic strategy is to show that the ``gauge'' transformation $\phi\rightarrow \phi+\gamma$ results in a field symmetry. To this end we can leverage Noether's Theorem and write the resulting change in the Lagrangian density as the divergence of a vector, the Noether current, which can then be set equal to zero.  As a direct consequence, the polarization angle must obey $D\gamma/Dz=0$ that is, it does not change along the optical path.  This behavior has long known to be true among those in the field of optics, but we believe this to be the first mathematical justification obtained from Lagrangian mechanics.

Recall the free-space, paraxial Lagrangian for homogeneously polarized light
\begin{align}
\mathcal{L}(\rho,\phi,\gamma,\vec{x},z)&=-k_0\rho\partial_z \phi-\frac{\rho}{2}|\nabla_X\phi|^2-\frac{\nabla_X\rho\cdot\nabla_X\rho}{8\rho}
\label{eqn:lagrangeDens0}
\end{align}
from which taking the variations with respect to $\rho,~\phi$ yield, respectively, the field equations
\begin{subequations}
\begin{align}
\frac{\delta \mathcal{L}}{\delta\rho}&\rightarrow k_0\partial_z\phi+\frac{1}{2}\nabla_X\phi\cdot\nabla_X\phi=\frac{1}{2}\frac{\nabla_X^2\rho^{1/2}}{\rho^{1/2}},
\label{eqn:phase}
\end{align}
\begin{align}
\frac{\delta\mathcal{L}}{\delta\phi}&\rightarrow \partial_z\rho+\nabla_X\left[\rho\nabla_X\phi\right]=0.
\label{eqn:intensity}
\end{align}
\label{eqn:Euler-Lagrange}
\end{subequations}
These equations are solved by the same electric field (amplitude and phase) that that are produced by the Huygen's Fresnel integral, that is, they solve the scalar, paraxial wave equation.  Now perform the transformation $\phi\rightarrow\phi+\epsilon\gamma$ where $\epsilon$ is small.  

The transformed Lagrangian density becomes
 \begin{align} \mathcal{L}'&=-k_0\rho\left(\partial_z\phi+\epsilon\partial_z\gamma\right)-\frac{\rho}{2}|\nabla_X\phi|^2-\epsilon \rho\nabla_X \phi\cdot\nabla_X\gamma \nonumber \\
     &~~~~~-\epsilon^2\frac{\rho}{2}|\nabla_X\gamma|^2-\frac{\nabla_X\rho\cdot\nabla_X\rho}{8\rho}\nonumber \\ 
     &=-\epsilon k_0\rho\frac{D\gamma}{Dz}-k_0\rho\partial_z\phi-\frac{\rho}{2}|\nabla_X\phi|^2\nonumber \\
     &~~~~~-\epsilon^2\frac{\rho}{2}|\nabla_X\gamma|^2-\frac{\nabla_X\rho\cdot\nabla_X\rho}{8\rho}
     \label{eqn:bigeikonal2}
 \end{align}
where we have used our familiar definition $\vec{v}\equiv k_0^{-1}\nabla_X\phi$ and hence the simplification
\begin{align}    \epsilon\rho\nabla_X\phi\cdot\nabla_X\gamma&=\epsilon k_0\rho (\vec{v}\cdot\nabla_X)\gamma.
\end{align}
Combining terms of order $\epsilon$ we see two additional terms in (\ref{eqn:bigeikonal2}),
\begin{align}
\delta\mathcal{L}=-\epsilon k_0 \rho \frac{D \gamma}{Dz}-\epsilon^2\frac{\rho}{2}|\nabla_X\gamma|^2.
\label{eqn:terms}
\end{align}

A symmetry in the current context is defined as those transformations that leave the action, $\mathcal{A}=\int \mathcal{L}d\vec{x}dz$, unchanged to leading order in $\epsilon$ \cite{Karatas:90,Brown:04}, that is,
\begin{align}
    \delta \mathcal{A}=O(\epsilon^2).
    \label{eqn:symmetry}
\end{align}
This will be the case if the variation in the Lagrangian density induced by the transformation (\ref{eqn:terms}) can be written as the divergence of a vector field (any two Lagrangian densities are equal up to such a term) \cite{Brown:04}. 


To find such a vector field, we can leverage Noether's Theorem.  When the Euler-Lagrange field equations are satisfied (i.e. when Eqns. \ref{eqn:Euler-Lagrange} hold), the Noether ``current'' associated to our transformation is \cite{Karatas:90,Brading:00}
\begin{align}
J_i&=\gamma \left\{\frac{\partial\mathcal{L}}{\partial (\partial_{x_i}\phi)}\right\},~i=1\cdots 3\nonumber \\
&=\gamma \{-k_0\rho,-\rho\partial_x \phi,-\rho\partial_y\phi\}\nonumber \\
&=\left\{-k_0\gamma\rho,-\gamma\rho\nabla_X\phi\right\}
\end{align}
where we have chosen the $i=1\cdots 3$ coordinates to be $x_1\equiv z,~x_2\equiv x, x_3\equiv y$.  Noether's Theorem then states that $\nabla\cdot{\bf J}=0$ and so
\begin{align}
\partial_z(\rho\gamma)+\nabla_X\cdot\left[\rho\gamma \vec{v}\right]&=0
\label{eqn:Noether}
\end{align}
which identifies the Noether charge
\begin{align}
    Q=\int_{X} \rho\gamma~d\vec{x}
    \label{eqn:charge}
\end{align}
as a conserved quantity.  Expanding (\ref{eqn:Noether}) we have that
\begin{align}
    \partial_z(\rho\gamma)+\nabla_X\cdot\left[\rho\gamma\vec{v}\right]&=0\nonumber \\
    \partial_z(\rho\gamma)+\rho\vec{v}\cdot\nabla_X\gamma+\rho\gamma\nabla_X\cdot\vec{v}+\gamma\vec{v}\cdot\nabla_X\rho&=0\nonumber \\
    \rho(\partial_z\gamma+\vec{v}\cdot\nabla_X\gamma)+\gamma(\partial_z\rho+\nabla_X\cdot\left[\rho\vec{v}\right])&=0\nonumber \\
    \rho\frac{D\gamma}{Dz}+\gamma(\partial_z\rho+\nabla_X\cdot\left[\rho\vec{v}\right])&=0.
    \label{eqn:derivation}
\end{align}
Since Eqn. (\ref{eqn:intensity}) holds we conclude
\begin{align}
    \frac{D\gamma}{Dz}&=0.
    \label{eqn:identity}
\end{align}
Thus, Noether's Theorem tells us that for a paraxial optical system that conserves intensity, the polarization angle does not change on propagation.  Moreover, we see that Eqn. (\ref{eqn:symmetry}) holds, hence to leading order the transformation $\phi\rightarrow \phi+\gamma$ is a symmetry. 

Of course, in our full model we retain the ``symmetry breaking term'' (second term in Eqn. \ref{eqn:terms} which is essential for demonstrating conservation of momentum i.e. writing Eqns. \ref{eqn:Cauchy3}).  However, the inclusion of the symmetry breaking term does not change the validity of (\ref{eqn:identity}) as one of our equations of motion in the transport model of beam propagation.\\*[1in]

\fi 
\bibliography{main}  
\end{document}